\documentclass[conference]{IEEEtran}
\IEEEoverridecommandlockouts
\usepackage[utf8]{inputenc}
\usepackage{cite}
\usepackage{amsmath,amssymb,amsfonts}
\usepackage{algorithmic}
\usepackage{subfig}
\usepackage{float}
\usepackage{graphicx}
\usepackage{textcomp}
\usepackage{xcolor}
\usepackage{tabularx} 
\usepackage{tabularray}
\usepackage{pifont}
\usepackage{multirow}
\usepackage{wrapfig}
\usepackage{soul}
\usepackage{blkarray} 
\usepackage[ruled,vlined]{algorithm2e} 
\usepackage{capt-of}
\usepackage{caption}
\usepackage{array}
\usepackage{hyperref}

\usepackage[export]{adjustbox}%

\AtBeginDocument{%
  \providecommand\BibTeX{{%
    \normalfont B\kern-0.5em{\scshape i\kern-0.25em b}\kern-0.8em\TeX}}}
    
\usepackage{tikz}

\begin{document}

 \makeatletter
    \newcommand{\linebreakand}{%
      \end{@IEEEauthorhalign}
      \hfill\mbox{}\par
      \mbox{}\hfill\begin{@IEEEauthorhalign}
    }
    \makeatother

\title{\huge REDACTOR: eFPGA Redaction for DNN Accelerator Security}

\author{\IEEEauthorblockN{Yazan Baddour}
    \IEEEauthorblockA{\textit{Electrical Engineering} \\
    \textit{California State Uni. Long Beach}\\
    Long Beach, CA, USA\\
    Yazan.Baddour01@student.csulb.edu}
    \and
    \IEEEauthorblockN{Ava Hedayatipour}
    \IEEEauthorblockA{\textit{Electrical Engineering} \\
    \textit{California State Uni. Long Beach}\\
    Long Beach, CA, USA\\
    Ava.Hedayatipour@csulb.edu}
    \and
    \IEEEauthorblockN{Amin Rezaei}
    \IEEEauthorblockA{\textit{Computer Engineering \& Computer Science} \\
    \textit{California State Uni. Long Beach}\\
    Long Beach, CA, USA\\
    Amin.Rezaei@csulb.edu}
}

\maketitle 

\begin{abstract}
With the ever-increasing integration of artificial intelligence into daily life and the growing importance of well-trained models, the security of hardware accelerators supporting Deep Neural Networks (DNNs) has become paramount. As a promising solution to prevent hardware intellectual property theft, eFPGA redaction has emerged. This technique selectively conceals critical components of the design, allowing authorized users to restore functionality post-fabrication by inserting the correct bitstream. In this paper, we explore the redaction of DNN accelerators using eFPGAs, from specification to physical design implementation. Specifically, we investigate the selection of critical DNN modules for redaction using both regular and fracturable look-up tables. We perform synthesis, timing verification, and place \& route on redacted DNN accelerators. Furthermore, we evaluate the overhead of incorporating eFPGAs into DNN accelerators in terms of power, area, and delay, finding it reasonable given the security benefits.
\end{abstract}


\section{Introduction}
\label{Introduction}
Deep Neural Network (DNN) technologies continue to evolve, propelled by the quest for enhanced accuracy. Achieving superior accuracy in DNNs necessitates high-performance computing resources, such as hardware accelerators. As accuracy and throughput become increasingly critical, the importance of security for DNN models and accelerators also escalates.

Attackers have the capability to reverse engineer the Intellectual Property (IP) of hardware accelerators, enabling them to create unauthorized substitute models. Additionally, they can tamper with the weights of the DNN, resulting in alterations to accuracy and misclassification of inputs \cite{MITTAL2021102163}. To address these security concerns, logic locking \cite{6241494, 9_Sequential_logic_encryption, DKLock, 6_Distributed_logic_encryption, 10_CoLA, Rezaei:KGL} has emerged using various types of key gates in which only authorized users can unlock the circuit. However, logic locking faces challenges from SAT-based attacks \cite{7140252, 8203759, 8942049, 10069713, 8714924, 7_Global_attack} that exploit an activated Integrated Circuit (IC) as a reference point along with the leaked locked netlist, allowing them to extract the correct key. In response to these threats, programmable fabrics exemplified by embedded Field-Programmable Gate Arrays (eFPGAs) offer robust defenses against reverse engineering and hardware IP theft \cite{8714856, 9371513, 9643548, 9473910}. The configuration bitstream remains accessible solely to the designer or authorized user, even if an untrusted foundry manufactures the design. The challenges in eFPGA redaction encompass identifying which modules within the IP should be redacted and minimizing the delay and area overheads associated with replacing these specific modules with eFPGAs.

To the best of our knowledge, \textbf{there is a gap in the existing literature on how eFPGA redaction affects the security of DNN accelerators.} Specifically, there is limited exploration of the implications on timing, area, and power overheads when integrating eFPGA to replace a segment of a hardware accelerator. The contributions of this paper are as follows:

\begin{itemize}
    \item [$\bullet$] Proposing an eFPGA redaction flow to hide sensitive components of a DNN accelerator with low overhead;
    \item [$\bullet$] Providing guidance on selecting a critical module to be replaced with regular or fracturable LUTs;
    \item [$\bullet$] Evaluating the overhead of the proposed redaction method in terms of area, power, and delay as well as the security against oracle-guided SAT-based attacks.
\end{itemize}

The rest of the paper is organized as follows: Section \ref{background} discusses the background, preliminaries, and related works. Section \ref{contribution} proposes our contributions to redact critical IPs of a DNN accelerator via eFPGAs, followed by verification, synthesis, and place \& route steps. Section \ref{experiment} depicts the comprehensive experimental results on the overhead and security of the redacted accelerator. Finally, conclusions are given in Section \ref{conclusion}.

\begin{figure*}[]
\centering
\subfloat[ 4x4 fabric ]
{
    \includegraphics[width=0.68\columnwidth]{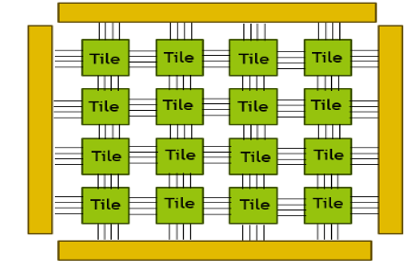}
    \label{4x4 fabric}
} \hspace{0.5em}
\subfloat[Tile components]
{
    \includegraphics[width=0.51\columnwidth]{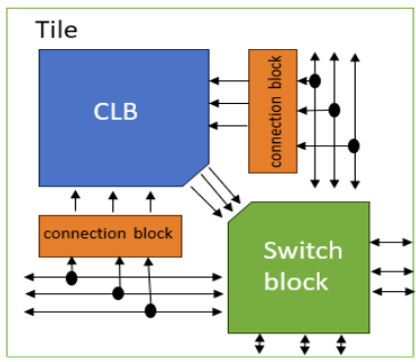}
    \label{tile}
} \hspace{0.5em}
\subfloat[BLE components]
{
    \includegraphics[width=0.66\columnwidth]{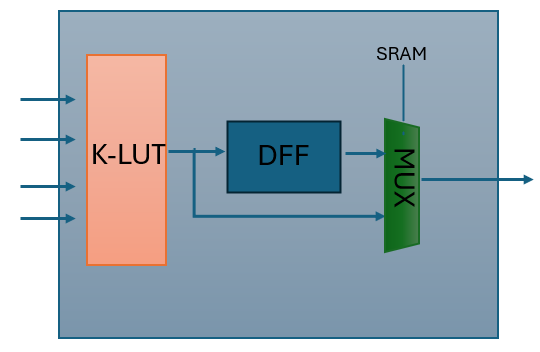}
    \label{BLE components}
}
\caption{eFPGA architecture}
\label{eFPGA Architecture}
\end{figure*}

\section{Background and Preliminaries} 
\label{background}
In this section, we discuss the background on DNNs, accelerators, eFPGAs, and existing work on securing hardware accelerators.

\subsection{Deep Neural Networks}
DNNs are employed in various tasks such as image classification \cite{5206848} and speech recognition \cite{7913606}. The input layer of DNNs receives inputs, and outputs are generated through weighted sums followed by non-linear activation functions. These outputs then proceed to subsequent layers. Initially set randomly, weights and biases are adjusted during the training phase to minimize disparities between expected and actual outputs. 

A Convolutional Neural Network (CNN) is a type of DNN comprising four main layers: convolutional, normalization, pooling, and fully connected layers. In the convolutional layer, inputs and outputs are processed as 2D or 3D arrays, with dimensions represented by height, width, and number of channels. The convolutional layer uses 3D filters to calculate the inner product with sub-arrays of the input, moving a sliding window of the filter size across the input with a fixed stride. Normalization layers adjust activation levels within each feature map, aiding in stabilizing the learning process by keeping inputs within a reasonable range. Pooling layers in CNNs down-sample input feature maps to create lower-resolution versions, retaining essential information while excluding irrelevant details. Operations such as average pooling and max pooling are applied independently to each feature map, reducing dimensions while maintaining the number of channels. Fully connected layers in CNNs determine class scores by processing outputs from preceding layers. Activation functions like Rectified Linear Unit (ReLU) introduce non-linearity.

\subsection{Hardware Accelerators} 
DNN models demand substantial computational resources, with tasks such as image classification requiring millions of weights and over half a billion Multiply And Accumulate (MAC) operations, necessitating specialized accelerators for real-time execution. DNN architectures require Arithmetic Logical Units (ALUs) \cite{10.1145/2749469.2750389, 10.1145/2694344.2694358} that focus on MAC units for computational resources and flexibility, dataflow optimization \cite{8358031, 8192478} to minimize energy consumption, and sparsity \cite{10.1109/ISCA.2016.11} to skip zero multiplications and reduce power consumption. 

Eyeriss \cite{7551407} is a Row-Stationary (RS) dataflow-based accelerator designed to minimize data movement energy in spatial architectures. In the RS dataflow, a row of operands is stored in the Register File (RF). Eyeriss employs diagonal connections of Processing Elements (PEs) for input reuse and vertical accumulation of partial sums. Each PE includes local registers for storing at least one row of weights and activations, along with a MAC unit and controller responsible for the temporal reuse of MAC units in 1D convolution. A significant portion of the RF is allocated to weights, and input vectors are reused to calculate partial sums for multiple output feature maps.

\subsection{Embedded Field Programmable Gate Arrays}
\label{Sec:eFPGAs}
Modern eFPGA architectures adopt a tile-based structure, with each tile containing configurable logic resources. Surrounding these tiles are the I/O blocks, which facilitate communication between the eFPGA and external devices or systems. Each tile within the eFPGA architecture comprises two fundamental elements that collectively enable its programmability and functionality. The first component is the Configurable Logic Block (CLB), which serves as the primary unit for implementing logic functions and user-defined designs. The second component is programmable routing, which facilitates the interconnection of various CLBs and enables the flow of data between them. Within each CLB, there is a Basic Logic Element (BLE) that incorporates a $K$-input Look-Up Table (LUT) capable of mapping a $K$-input single-output Boolean function, alongside a flip-flop and a 2-1 routing multiplexer used to toggle between sequential and combinational logic. 

Fig. \ref{eFPGA Architecture} provides an example of a 4x4 eFPGA architecture, the components inside each tile, and a visual representation of a basic BLE setup with a 4-input LUT. Studies have identified the optimal LUT size for balancing area and delay to be between 4 to 6, with 6-LUTs exhibiting superior performance and 4-LUTs occupying the smallest area \cite{121549}. To address these trade-offs and enhance LUT utilization, a refined version known as fracturable LUTs (FLUTs) has been introduced. FLUTs have more than one mode of operation: they can function as a $K$-input LUT or be fractured into smaller $K$-1 LUTs, with shared inputs \cite{6132691}. Within the CLB, BLEs can be organized into logic clusters to optimize eFPGA speed and area efficiency. The ideal cluster size ($N$) is typically determined to be between 4 and 10 \cite{606687}. Crossbar routing interconnects the inputs and outputs of CLBs and BLEs using a series of programmable multiplexers, ensuring connectivity between all BLEs and every CLB pin. Switch Blocks (SBs) and Connection Blocks (CBs) serve as the global routing that connects CLBs together.

\subsection{Related Works}
With the recent boom in Artificial Intelligence (AI), developing new approaches to safeguard the DNN models and accelerators is of the utmost importance. In \cite{10.5555/3199700.3199718}, fault injection attacks are examined that are capable of inducing misclassification in DNN by altering the bias in the output layer to favor the desired adversarial class. As a countermeasure in \cite{10.1145/3359789.3359831} the fault-sensitivity of individual neurons within a given DNN is measured by effectively leveraging both external and internal redundancy within DNN models to balance system robustness against hardware overhead. 

In addition, a Hardware Protected Neural Network (HPNN) framework is proposed \cite{9218651} to safeguard DNNs from attackers with extensive knowledge. It conceals the weight space through a confidential HPNN key, controlling each neuron's functionality. In \cite{10143985} a hardware key is introduced to secure the accelerator such that a wrong key increases memory access, inference time, and energy consumption, making it unsuitable for inference. This method requires modifying the activation function to prevent bypassing the hardware key controlled block. Moreover, they use a model key to obscure the model without the need of retraining the model. Although this approach resulted in decreased accuracy and higher memory access when incorrect keys were applied, it necessitates modifications in the DNN accelerator. 

Initial logic locking solutions utilize XOR-based and MUX-based mechanisms \cite{EndingPiracy, 1fault_analysis}. However, the oracle-guided SAT attack \cite{7140252} exposes vulnerabilities in these methods, prompting the development of more robust techniques \cite{AntiSAT, SARLock, SFLL, cryptoeprint:2019/139, 6_Distributed_logic_encryption, 9_Sequential_logic_encryption, 10_CoLA, MLOverheadAnalysisofLogicLocking, 5_CyclicL}, which increase the time complexity of SAT attacks.

Furthermore, a soft embedded FPGA redaction method is introduced \cite{9473910} to hide critical IP functionality and routing within RTL designs. A critical IP in the Verilog file is identified and synthesized using Yosys \cite{Wolf2013YosysAFV}, then place \& route is performed to determine the smallest eFPGA fabric needed. The resulting fabric, devoid of critical IP information, is loaded with a bitstream to maintain functionality. In \cite{10225483} the effects of varying parameters ($K$ and $N$) on area, power, delay, and security of eFPGA architectures are compared. Increasing $K$ influences CLB inputs, impacting LUT sizes and routing, while increasing $N$ adds BLEs, affecting area and routing complexities.  Findings challenge the assumption that fabric size directly correlates with security strength. 

In \cite{10069713}, the assumption that eFPGA-based designs are inherently secure against oracle-guided attacks has been challenged. The researchers conducted two attacks: CycSAT \cite{8203759}, which aims to break cycles within the circuit, and IcySAT \cite{8942049}. IcySAT took longer due to the unrolling process. However, CycSAT struggles to break hard loops, leaving at least one loop intact and causing the SAT solver to repeat iterations. To address this, the paper developed a two-phase attack called Break \& Unroll. The first phase breaks cycles sequentially, creating a new non-cyclic constraint. If hard cycles persist, the second phase unrolls the circuit, duplicating gates and breaking feedback connections to prevent infinite loops.

\begin{figure*}[!t]
   \includegraphics[width=\textwidth]{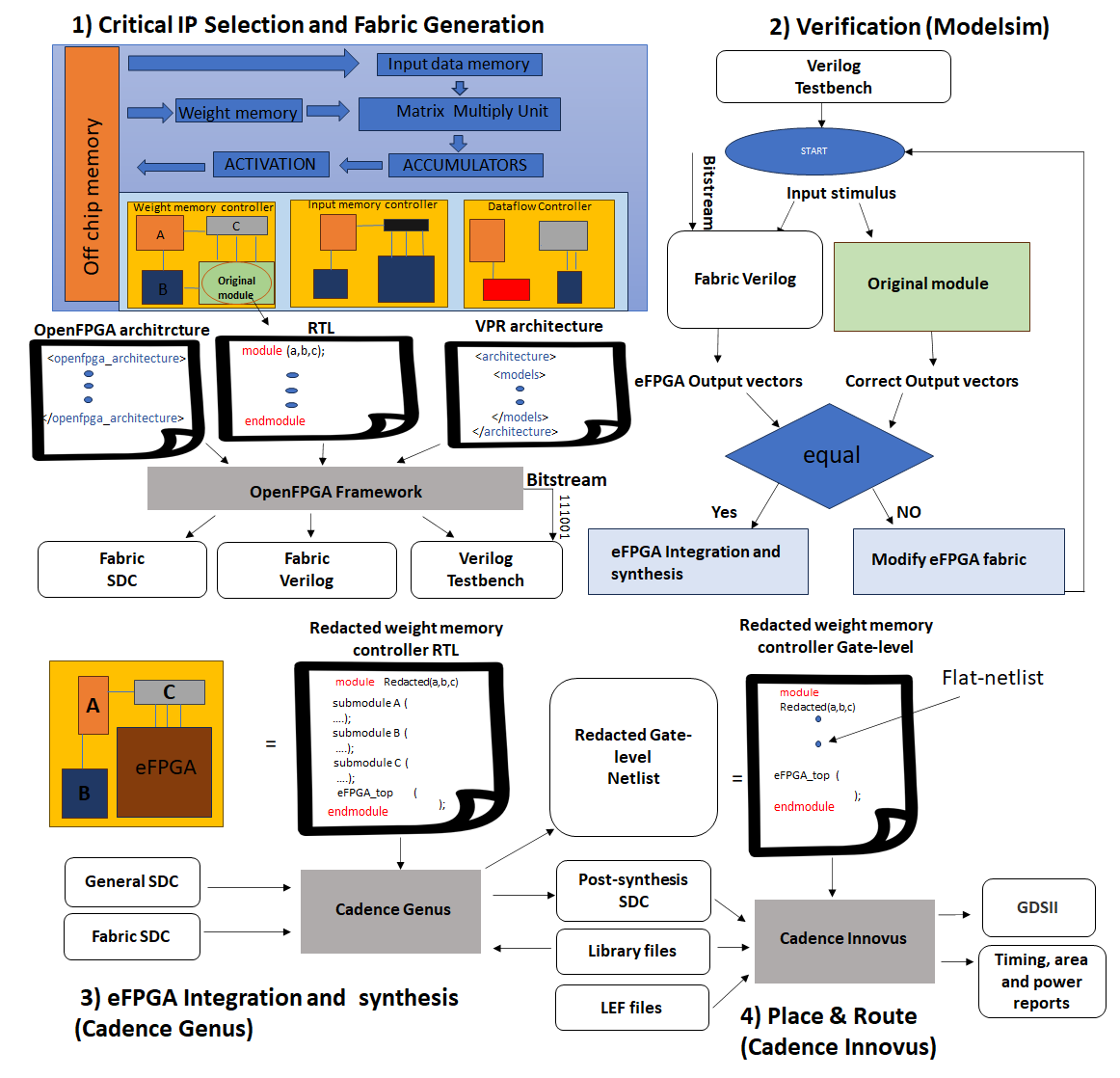}
   \caption{The proposed eFPGA redaction flow for DNN accelerators}
  \label{MAIN}
\end{figure*}

\section{REDACTOR} 
\label{contribution}
In this section, we propose \textbf{REDACTOR}, an eFPGA \textbf{RED}action framework for CNN \textbf{AC}celera\textbf{TOR}s shown in Fig. \ref{MAIN}. Without loss of generality, we utilize the accelerator outlined in \cite{githubGitHub8krisvCNNACCELERATOR}, which is a modification of the Broadcast, Stay, Migration (BSM) dataflow initially proposed in \cite{8374840}. Although the procedure is independent of the chosen accelerator, this adaptation is considered energy-efficient as it effectively minimizes redundant access to off-chip memory.

\subsection{Critical IP Selection and Fabric Generation}
The goal of module selection is to choose a module that plays a crucial role in the overall functionality of the accelerator. In this regard, three options can be considered: 

\begin{itemize}
\item Opting for a module that significantly impacts the outputs proves beneficial for causing output corruption when an unauthorized user loads the wrong bitstream.
\item Selecting a module that causes error propagation throughout the system can amplify the impact of using incorrect bitstreams and decrease the accuracy of the DNN model.
\item Choosing a module that can be implemented using a complex fabric with a large unroll factor can be beneficial for securing against existing attacks that target eFPGAs while ensuring it remains within the power, area, and delay budget.
\end{itemize}

The first critical IP selected is the On-Chip Weight Memory Controller (OWMC), which has 12 outputs, some of which directly and indirectly impact the loading of DNN weights from the weight memory to the data flow block comprising numerous Processing Elements (PEs). Protecting these weights ensures that sensitive information remains secure from unauthorized access or corruption. Subsequently, we iterate through the redaction process to redact several IPs in the accelerator using eFPGAs of varying sizes and architectures. These IPs include the Multiplexers Dataflow Controller (MUXDC), the On-Chip Memory Dataflow Controller (OMDC), and the Processing Elements Dataflow Controller (PEDC). Table \ref{Critical IPs redacted in this study} displays the characteristics of each critical IP. 

\begin{table}[!t]
     \footnotesize
      \centering
      \caption{Critical IPs redacted} 
      \begin{tblr}  {
        colspec = {|c|c|c|m{4.4cm}|},
      }
        \hline
        
          Critical & \# of &  \# of & Description  \\ 
          IP & Modules &  I/Os&  \\
        \hline
        \hline

          OWMC & 5 & 20& Controls the flow of DNN weights\\
        \hline
          MUXDC & 4 & 15& Controls dataflow between PEs\\
        \hline
          OMDC & 20  & 26& Controls the convolution process\\

        \hline
          PEDC & 1 & 5 & Sets/resets output register of PEs\\
                 \hline
   
      \end{tblr}

      \label{Critical IPs redacted in this study}
      
\end{table}

To generate the eFPGA fabrics, we use OpenFPGA \cite{9098028}. We input three files into OpenFPGA: the benchmark file, which contains the critical RTL Verilog portion of the selected module for redaction, and two architecture files. One is the OpenFPGA architecture file \cite{8576622}, and the other is the VPR architecture file \cite{10.1145/1950413.1950457}. Both files are in Extensible Markup Language (XML) format and together specify the architecture of the eFPGA. 

We select a fabric architecture to achieve 100\% block utilization and over 90\% I/O utilization. This is accomplished by manually adjusting the number of BLEs ($N$) per CLB and the number of I/O pins per tile until the target utilization is reached, while maintaining a constant 4-input LUT/FLUT ($K$=4). The number of CLB inputs can be determined using the following formula, known to provide favorable Power, Performance, and Area (PPA) characteristics \cite{1281800}.

\begin{equation}
    I = \frac{K(N+1)}{2}
\end{equation}

Next, we acquire the timing constraints (i.e., SDC files) for the fabric intended for use in the physical design phase. Additionally, we obtain the Verilog files for the fabric, along with a testbench designed to verify the functionality of the fabric and the bitstream necessary to program the fabric to operate as the chosen module for redaction. Table \ref{eFPGA-parameters} presents the eFPGA parameters used.

\begin{table}[!t]
    \footnotesize
      \centering
      \caption{eFPGA parameters used} 
      \begin{tblr}{
        colspec = {|c|c|m{5.5cm}|},
      }
        \hline
          Parameter & Value & Description\\ 
        \hline
        \hline

          K & 4 & Input size of a LUT/FLUT\\
        \hline
          N & [1,9] & Number of BLEs per CLB\\
        \hline
         W  & Auto & Number of routing tracks in a channel\\
        \hline
         Fc$_{in}$  & 0.15 & Fraction of the routing tracks that a CLB input can connect \\

        \hline
         Fc$_{out}$  & 0.1 & Fraction of the routing tracks that a CLB output can connect \\
                 \hline
         Fs  & 3 & Number of connections per incoming routing track in a switch block  \\
                 \hline
         L  & 4 & The length of routing track (number of CLBs spanned) \\
                 \hline

      \end{tblr}
    \label{eFPGA-parameters}
\end{table}

\subsection{Verification}
We utilize ModelSim to simulate the functionality of the fabric. Both the original design and the eFPGA fabric receive shared input stimuli, and the correct bitstream is loaded into the fabric for evaluation. Subsequently, the output vectors of the eFPGA and the original design are compared. When the correct bitstream is applied, the output vectors should match, ensuring consistency and functionality between the eFPGA and the original design. In Fig. \ref{eFPGA and original module output vectors}, the waveforms depict the output vectors. The eFPGA output signals are visualized in green, while the original module's signals are represented in yellow. The verification process is repeated for the synthesized fabric to verify functionality after synthesis.

\begin{figure}[!b]
  \centering
   \includegraphics[width=\columnwidth, left]{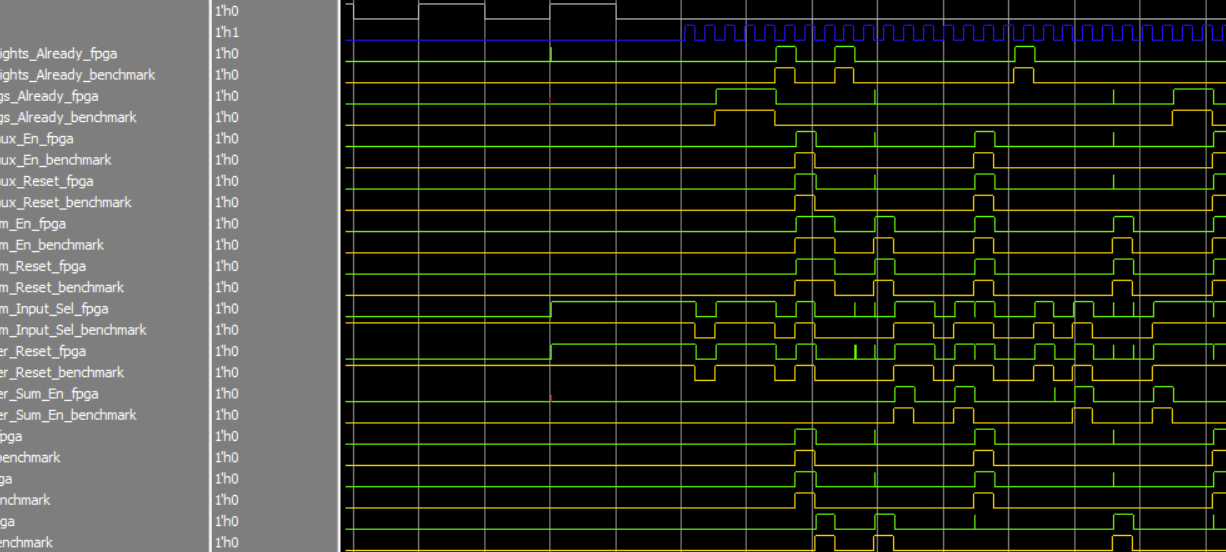}
   \caption{eFPGA fabric and original module output vectors}
  \label{eFPGA and original module output vectors}
\end{figure} 

\subsection{eFPGA Integration and Synthesis}
We utilize Cadence Genus for synthesizing each redacted module. The redacted module serves as the top module in Cadence Genus for synthesis, where we replace the critical Verilog portion with the eFPGA at the RTL level. To map the Verilog netlist to a gate-level netlist, we employ the standard cell library file available on the Cadence website (i.e., slow\_vdd1v2\_basicCells.lib), which is a 45nm technology library characterized by slower operating speeds or slower process corners. 

First, we read the Verilog and library files and elaborate the top module. Then, we analyze all the SDC files generated by OpenFPGA and command Genus to report timing. At this stage, Genus addresses any combinational loops that legally exist within the eFPGA fabric by inserting a buffer from the technology library onto the feedback loop (i.e., cdn\_loop\_breaker\_cell). Additionally, it disables the timing arc from the input to the output, effectively breaking the timing loop. Fig. \ref{cdn_loop_break.png} shows an example of a loop that is broken. The buffer highlighted in yellow represents the feedback loop.

We choose to use a flattened netlist to simplify the place and route stage. However, during the process of flattening the netlists, Cadence Genus modifies the names of hierarchical instances, which results in the disregard of some initial SDC constraints. To address this issue, we instruct Genus to parse an additional SDC file, which contains instructions to disable timing for specific paths using the flattened names.

Finally, optimization is performed on the flattened netlist, where we instruct the tool to attempt optimization on all paths with negative slack, including the critical path. Reports are then generated to provide details on area, power, and timing. Additionally, the gate-level netlist Verilog file and a single SDC file containing all constraints are generated. These two files serve as inputs for the subsequent place \& route stage. Genus does not incorporate the constraints necessary to break the timing loops in the final SDC file. Therefore, we manually modify the file to include these constraints. 

\begin{figure}[!t]
  \centering
   \includegraphics[width=\columnwidth, left]{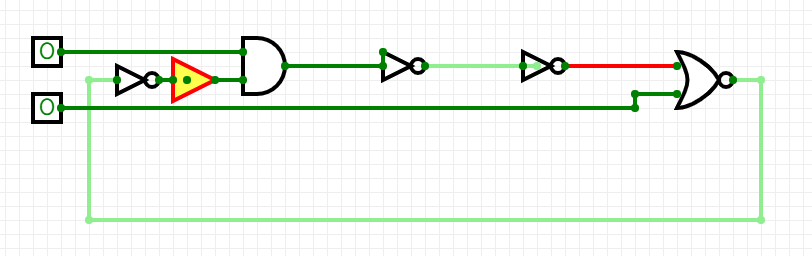}
   \caption{Loop breaker cell example}
  \label{cdn_loop_break.png}
\end{figure}

\begin{figure}[!b]
  \centering
   \includegraphics[width=\columnwidth, left]{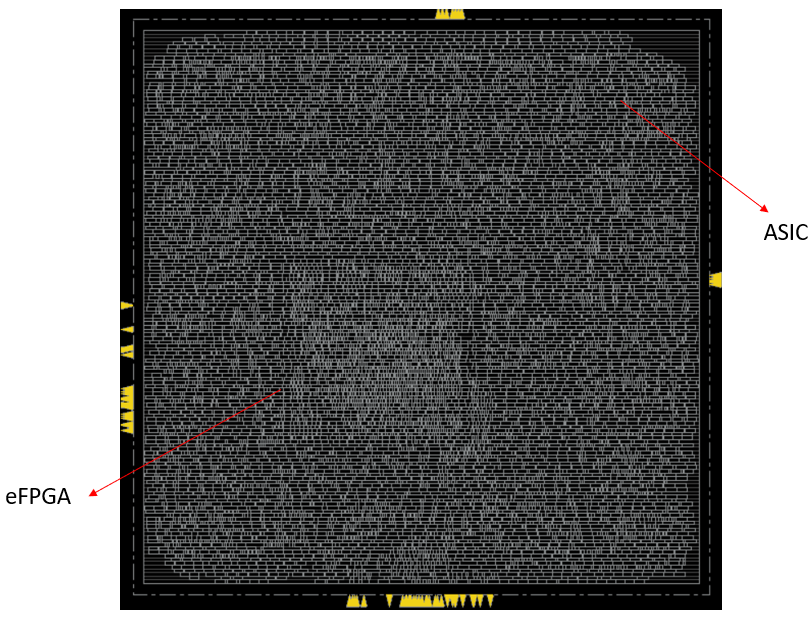}
   \caption{Redacted module layout without metal layers}
  \label{layout}
\end{figure}

\subsection{Place \& Route}
The place \& route stage in IC design is a critical process where logic gates, flip-flops, and other components undergo positioning (i.e., placement) and interconnection (i.e., routing). We utilize Cadence Innovus for the place \& route of the redacted modules. Cadence Innovus takes inputs including the gate-level netlist and the SDC constraint file generated by Cadence Genus, along with technology files such as timing library files (.lib) and library exchange format (.lef) files.

The process begins with specifying the floor plan, followed by power planning, which involves inserting VDD and VSS rings, and power and ground stripes to connect them to standard cells. Standard cells are then placed within the specified floor plan. Additionally, we instruct Innovus to place the design with I/O pins, eliminating the need for manual I/O file creation. Subsequently, pre-Clock Tree Synthesis (CTS) optimization is performed to meet all timing constraints before CTS. We then perform CTS, followed by post-CTS optimization to further refine the design to meet timing constraints. Finally, routing and post-route optimization are carried out to complete the process, and timing, area, and power reports are generated. Fig. \ref{layout} shows the layout of the redacted module without metal layers. The eFPGA is noticeable in the middle, where it exhibits a slightly different pattern.

\begin{table}[!t]
    \footnotesize
      \centering
      \caption{eFPGA fabric characteristics} 
      \begin{tblr}{
        colspec = {|c|c|c|c|c|},
      }
        \hline
          eFPGA &Block &I/O  & Bitstream & Channel\\ 
        Fabric& Utilization & utilization & Size& Width\\
        \hline
        \hline

            2x2 K4N2 & 100\% & 94\% & 614 & 18\\
            \hline
     
             2x2 K4\_frac\_N1& 100\% & 94\% & 458 & 18\\
      
        \hline

            1x1 K4N6 & 100\% & 100\% & 440 & 26\\
            \hline
     
             1x1 K4\_frac\_N3& 100\% & 100\% & 256 & 18\\
        \hline        

             2x2 K4N4 & 100\% & 95\% & 1160 & 30\\
            \hline
     
             2x2 K4\_frac\_N3& 100\% & 95\% & 1059 & 30\\
        \hline

             1x1 K4N1 & 100\% & 100\% & 66 & 6\\
        \hline
     
             1x1 K4\_frac\_N1& 100\% & 100\% & 79 & 14\\
        \hline        
          \end{tblr}
    \label{eFPGA characteristics}
\end{table}

\section{Experimental Results} 
\label{experiment}
In this section, we evaluate the overhead and security of \textbf {REDACTOR}. The source codes, along with the created eFPGA fabrics, are available on our GitHub repository\footnote{https://github.com/cars-lab-repo/REDACTOR}. 

\begin{figure*}
\centering
\subfloat[ Redacted OWMC ]
{
    \includegraphics[width=0.33\textwidth]{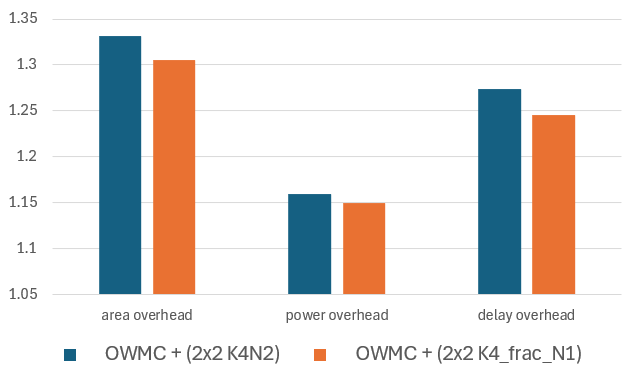}
    \label{overhead_a}
}
\subfloat[ Redacted MUXDC ]
{
    \includegraphics[width=0.33\textwidth]{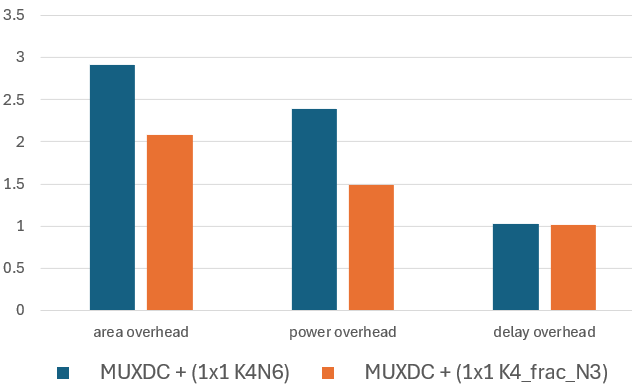}
    \label{overhead_b}
}
\subfloat[Redacted OMDC]
{
    \includegraphics[width=0.33\textwidth]{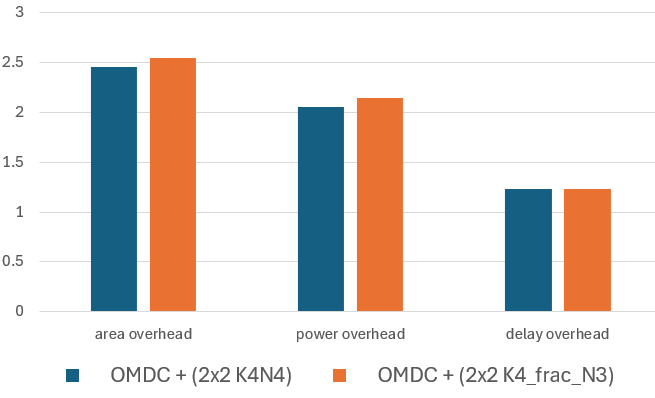}
    \label{overhead_c}
}

\caption{Normalized area, power, and delay overhead of LUT-based vs. FLUT-based eFPGAs redacted modules}
\label{over_heads}
\end{figure*}

\subsection{Overhead Analysis}
As mentioned earlier, our goal is to find the simplest fabric using the parameters shown in Table \ref{eFPGA-parameters}. We achieve this by increasing or decreasing the number of BLEs in each CLB and I/O pins in I/O tiles. The characteristics of the eFPGA fabrics utilized in the experiments are summarized in Table \ref{eFPGA characteristics}. Fig. \ref{over_heads} compares the area, power, and delay overheads of LUT-based eFPGA and FLUT-based eFPGA for the redacted modules normalized according to the original designs.

\textbf{OWMC:} We redact the OWMC IP using two fabrics: The first one is a LUT-based fabric with 2 BLEs/CLB (2x2 K4N2). The second one is a FLUT-based fabric with 1 BLE/CLB (2x2 K4\_frac\_N1). Despite the expected larger bitstream for the FLUT-based fabric, it actually has a smaller size because of fewer BLEs/CLB. Both fabrics fully utilize blocks and I/O but have different bitstream sizes: 614 for the LUT-based and 458 for the FLUT-based. Despite FLUT's anticipated impact on channel width, it remains the same at 18 for both fabrics due to fewer BLEs. The LUT-based fabric introduces higher area, power, and delay overheads (33\%, 16\%, and 27\% respectively) compared to the FLUT-based fabric (30\%, 15\%, and 24.5\% respectively). \textit{Opting for FLUT-based is preferable when integrating eFPGA for the OWMC IP.}

\textbf{MUXDC:} We test the MUXDC IP using two fabric configurations: a LUT-based with 6 BLEs/CLB (1x1 K4N6) and a FLUT-based with 3 BLEs/CLB using FLUT (1x1 K4\_frac\_N3). Both fabrics achieve full block and I/O utilization. The LUT-based has a larger bitstream size of 440 compared to the FLUT-based, which has 256 bits. Routing widths differ, with the LUT-based at 26 and the FLUT-based at 18 tracks per channel. Due to this variance, the area difference between the fabrics is notable. The FLUT-based shows significant improvement in power consumption and a slightly better critical path delay. The first fabric introduces higher area, power, and delay overheads (192\%, 139\%, and 2\% respectively) compared to the second fabric (107\%, 49\%, and 1.6\% respectively). \textit{For the MUXDC IP, opting for FLUT proves preferable for redaction.}

\textbf{OMDC:} In addition, the OMDC IP is tested using two fabrics: a LUT-based (2x2 K4N4) and a FLUT-based (2x2 K4\_frac\_N3), both achieving full block utilization and 95\% I/O utilization. Despite the LUT-based fabric having a larger bitstream size of 1160 compared to the second fabric's 1059, the fabric utilizing regular LUTs shows slightly better performance in area, power, and delay, even though both fabrics routed with a channel width of 30. The LUT-based introduces an area overhead of 145\%, a power overhead of 105\%, and a critical path delay overhead of 23\%, while the FLUT-based introduces an area overhead of 154\%, a power overhead of 114\%, and a critical path delay overhead of 23.6\%. \textit{For the OMDC IP, choosing a regular LUT is preferable.}

\textbf{PEDC:} Finally, the PEDC IP is evaluated with two fabrics: a basic LUT-based with one CLB and one BLE (1x1 K4N1), and an identical fabric with FLUT instead (K4\_frac\_N1). Both fabrics utilize I/Os and resources fully. The LUT-based fabric needs 66 bits for programming, while the FLUT-based one requires 79. The LUT-based fabric has a minimum channel width of 6, while the FLUT-based one needs a wider channel width of 14 due to its more complex routing. Because the original PEDC IP is a very small finite state machine, both the LUT-based and FLUT-based fabrics shows huge area and power overhead as well as high delay compared with the original IP. \textit{In the case of PEDC IP, choosing the fabric with regular LUTs is preferable. However, it is not efficient to replace a small IP with eFPGA due to the overheads introduced.}

\begin{table}[!t]
    \footnotesize
      \centering
      \caption{Security results} 
      \begin{tblr}{
        colspec = {|c|c|c|c|c|},
      }
        \hline
               Fabric& Unroll & \# Clauses & Time (s) & Key Reported?\\ 
        \hline
        \hline

             2x2 K4N2 & 59 & 1610318 & 99 & yes\\
            \hline
     
             2x2 K4\_frac\_N1& 30  &559536  & 7 & yes\\
      
        \hline

             1x1 K4N6 & 36 & 875500 & 51 & yes\\
            \hline
     
             1x1 K4\_frac\_N3& 2 & 9291 & 0.26  & yes \\
        \hline        

             2x2 K4N4 & 112 & N/A & Time out & no\\
            \hline
     
             2x2 K4\_frac\_N3& 59  & 3299375 & 81 & yes\\

        \hline
             1x1 K4N1 & 5 & 4189 & 0.1 & yes\\
        \hline
     
            1x1 K4\_frac\_N1& 2 & 2384 & 0.08 & yes \\
        \hline        
      \end{tblr}
    \label{security_results}
\end{table}

\subsection{Security Analysis}
We conduct a separate synthesis of the eFPGA fabric, this time constraining Genus to utilize only basic logic cells. Using Python, we convert the Verilog gate-level netlist from Genus to a $.bench$ file. To establish an oracle, we program the fabric with the bitstream generated by OpenFPGA, assigning the outputs of the DFFs found in the scan chain with their corresponding bits. For creating a key-controlled netlist, we expose the outputs of the DFFs in the scan chain as key inputs. Then, we utilized the break and unroll attack found in \cite{10069713}, which is then fed to NEOS \cite{bitbucketBitbucket} tool for extracting the bitstream. 

The security results are presented in Table \ref{security_results}. We were able to extract the bitstream of all eFPGAs that used FLUTs in a shorter time compared to eFPGAs that used regular LUTs. This is because the LUT-based fabrics have a higher unroll factor. For instance, our 2x2 K4N4 fabric timed out after 6 hours of running the attack, which has an unroll factor of 112. To maintain the reduced overheads of FLUT-based eFPGA fabrics, we propose two methods for future research: 

\begin{itemize}
    \item [$\bullet$] Introducing more cycles within the eFPGA fabric, thereby increasing the unroll factor to the point where the attack times out. 
    \item [$\bullet$] Introducing non-unrollable cycles within the eFPGA fabric, which consist of oscillating and stateful cycles \cite{10069713} interlaced in a manner that ensures at least one cycle remains unbroken during a cyclic attack, causing it to fail by entering an infinite loop.
\end{itemize}

\section{Conclusion}
\label{conclusion}
In this paper, we explored the importance of securing DNN accelerators and proposed an approach for redacting critical IPs with eFPGAs, from specification to physical design. Specifically, we focused on evaluating the impact of eFPGA fabrics with high I/O and block utilization and assessed the integration of these nearly fully utilized fabrics with regular LUTs and FLUTs. 

While using FLUT-based eFPGAs can complicate routing, they offer an advantage by reducing the number of BLEs or CLBs, which generally translates to lower power, delay, and area overhead. As demonstrated by the experiments, the choice between LUT or FLUT depends on the specific IP to be redacted, and no general rule can be established. Additionally, we observed that FLUT-based eFPGAs have lower unroll factors, making it easier for attackers to extract the bitstream. Conversely, when we redacted the OWMC IP with a regular LUT-based eFPGA, the attack timed out, and we were able to maintain reasonable overhead while enhancing security.

For future research, systematically increasing the unrolling factor or adding non-unrollable cycles to eFPGA fabrics can be pursued to maintain low overhead while improving security.

\section*{Acknowledgment}
This material is based upon work supported by the National Science Foundation under Award No. 2245247.

\end{document}